\documentclass[conference]{IEEEtran}
\IEEEoverridecommandlockouts
\usepackage{cite}
\usepackage{amsmath,amssymb,amsfonts}
\usepackage{quantikz}
\usepackage{algorithmic}
\usepackage{graphicx}
\usepackage{textcomp}
\usepackage{xcolor}
\def\BibTeX{{\rm B\kern-.05em{\sc i\kern-.025em b}\kern-.08em
    T\kern-.1667em\lower.7ex\hbox{E}\kern-.125emX}}
\begin{document}

\title{Constrained Counterdiabatic Quantum Approximate Optimization Algorithm for Portfolio Optimization\\
}

\author{\IEEEauthorblockN{Jose Falla}
\IEEEauthorblockA{\textit{Department of Physics and Astronomy} \\
\textit{University of Delaware}\\
Newark, DE, USA \\
jfalla@udel.edu}
\and
\IEEEauthorblockN{Ilya Safro}
\IEEEauthorblockA{\textit{Department of Computer and Information Sciences} \\
\textit{Department of Physics and Astronomy} \\
\textit{University of Delaware}\\
Newark, DE, USA \\
isafro@udel.edu}
}

\maketitle

\begin{abstract}
We introduce a counterdiabatic (CD) extension of the Quantum Approximate Optimization Algorithm (QAOA) for constrained portfolio optimization. By incorporating approximate adiabatic gauge potentials generated from nested commutators of the Ising-type portfolio problem Hamiltonian and the Hamming weight-preserving XY mixer Hamiltonian into our variational ansatz, the resulting Constrained Counterdiabatic QAOA (CCD-QAOA) achieves improved optimization performance under realistic budget and risk constraints. Benchmarking against standard XY-mixer QAOA, Grover-mixer QAOA, and penalty-based QAOA formulations, our numerical simulations demonstrate that, for a fixed QAOA depth, our CCD-QAOA approach consistently results in better approximation ratios.\\
\noindent {\bf Reproducibility: Link and data will be available upon acceptance.}
\end{abstract}


\section{\label{sec:introduction}Introduction}
Variational quantum algorithms are among the most promising approaches for leveraging near-term and fault-tolerant quantum hardware to address classically challenging optimization problems. The Quantum Approximate Optimization Algorithm (QAOA) \cite{farhi_quantum_2014} has emerged as a particularly versatile framework, providing a controllable interpolation between digitized adiabatic evolution and shallow variational circuits. QAOA has been applied in such areas as finance \cite{herman2023quantum}, network science \cite{shaydulin2018network}, and transportation \cite{azad2022solving}, to mention just a few. Despite its conceptual simplicity and broad applicability, QAOA performance often degrades for constrained combinatorial optimization problems, where competing energy scales and small spectral gaps hinder efficient state preparation. 

Many practically relevant optimization tasks are intrinsically constrained. Portfolio optimization, in particular, provides a natural and structured example: risk-return tradeoffs yield dense quadratic interactions, while budget constraints restrict feasible configurations to a narrow solution subspace. Standard penalty-based encodings introduce large energy penalties that distort the spectrum and can suppress effective gaps \cite{tang_comparative_2024, saleem_approaches_2023, kerenidis_quantum_2019}, while constraint-preserving mixers require problem-specific circuit constructions \cite{brandhofer_benchmarking_2022, fuchs_constrained_2022, wang_x_2020}. In both cases, the variational landscape can become difficult to navigate (in particular, at low circuit depth).

From the perspective of adiabatic quantum computation, these challenges can be traced to diabatic transitions generated during interpolation between mixing and cost Hamiltonians \cite{amin_consistency_2009}. Counterdiabatic (CD) driving offers a principled mechanism for suppressing such transitions by augmenting the Hamiltonian with an adiabatic gauge potential (AGP). While exact counterdiabatic terms are generally nonlocal and experimentally inaccessible, variational approximations provide a systematic and hardware-compatible route to incorporating CD corrections into digitized algorithms \cite{guery-odelin_shortcuts_2019, cepaite_counterdiabatic_2023, hegade_digitized_2022}.

\noindent {\bf Our contribution: }
In this work, we develop a counterdiabatic extension of QAOA tailored to constrained optimization problems and demonstrate its performance in the context of portfolio optimization. Our approach constructs approximate gauge potentials variationally within a structured operator algebra generated by the cost and constraint-preserving mixer Hamiltonians. This framework provides (i) a systematic method for generating physically relevant counterdiabatic operators, (ii) a controlled truncation strategy, and (iii) insight into how constraint structure influences ansatz complexity in CD-QAOA.

We show that incorporating counterdiabatic generators into QAOA substantially improves approximation ratios at fixed circuit depth for instances of the portfolio optimization problem using Hamming weight-preserving XY mixers as generators of adiabatic gauge potentials. Beyond the specific application to portfolio optimization, our results 
illustrate a broader principle: constrained combinatorial optimization 
problems possess structured operator algebras that can be leveraged 
to construct efficient counterdiabatic ansätze. This connection between 
adiabatic gauge potentials and variational quantum circuits provides 
a unifying framework for enhancing QAOA performance in structured 
optimization landscapes.

\section{Background}
\subsection{The Quantum Approximate Optimization Algorithm}

The Quantum Approximate Optimization Algorithm provides a variational framework for approximately preparing the ground state of a problem Hamiltonian encoding a combinatorial optimization task \cite{farhi_quantum_2014}. Starting from an easy-to-prepare reference state, QAOA constructs a parametrized quantum circuit consisting of alternating applications of cost and mixing unitaries.

At its core, QAOA translates a classical cost function into a Hermitian operator known as the \emph{problem Hamiltonian} $H_C$, whose lowest-energy state encodes the desired solution. The algorithm proceeds by alternating unitary evolutions generated by $H_C$ and a complementary \emph{mixer Hamiltonian} $H_M$. The role of $H_M$ is to drive transitions between computational basis states, thereby enabling exploration of the solution space and helping to avoid entrapment in suboptimal configurations. Despite its elegant formulation, QAOA encounters several practical limitations. Chief among these is the difficulty of optimizing its variational parameters using classical routines, particularly as the circuit depth $p$ increases. The resulting optimization landscape is typically highly nonconvex, leading to significant computational overhead that can erode any potential quantum advantage~\cite{ZWCHL}. A well-known manifestation of this challenge is the presence of \emph{barren plateaus}, where gradients vanish exponentially with system size, rendering training ineffective~\cite{mcclean2018barren,LTWS}.

Specifically, for a circuit of depth $p$, the QAOA state is defined as

\begin{equation}
    \ket{\psi(\boldsymbol{\gamma},\boldsymbol{\beta})} =
    \prod_{k=1}^{p}
    e^{-i \beta_k H_M}
    e^{-i \gamma_k H_C}
    \ket{\psi_0},
\end{equation}
where $H_C$ is the cost Hamiltonian encoding the solution to the combinatorial optimization problem, $H_M$ is a mixing Hamiltonian, and $\ket{\psi_0}$ is the initial state, usually taken to be the uniform superposition of the computational basis state on $N$ qubits. The parameters $\boldsymbol{\gamma} = (\gamma_1,\ldots,\gamma_p)$ and $\boldsymbol{\beta} = (\beta_1,\ldots,\beta_p)$ are optimized classically to minimize the expectation value 

\begin{equation}
    \bra{\psi(\boldsymbol{\gamma},\boldsymbol{\beta})}
    H_C
    \ket{\psi(\boldsymbol{\gamma},\boldsymbol{\beta})}.
\end{equation}
The standard choice of mixer is the transverse-field Hamiltonian
\begin{equation}
    H_M = \sum_{i=1}^{N} X_i ,
\end{equation}
where $X_i$ is the Pauli-$X$ operator acting on qubit $i$. This operator generates transitions between computational basis 
states and enables exploration of the Hilbert space during the variational optimization.

The corresponding initial state is typically chosen as the uniform superposition

\begin{equation}
    \ket{\psi_0} = \ket{+}^{\otimes N},
\end{equation}
which is the ground state of $H_M$. In the limit $p\to\infty$ with small steps, QAOA reproduces continuous adiabatic evolution \cite{farhi_quantum_2014}.

QAOA can be viewed as a Trotterized approximation to adiabatic 
evolution between the mixer and cost Hamiltonians. Consider the 
time-dependent Hamiltonian

\begin{equation}
    H(s) = (1-s) H_M + s H_C ,
\end{equation}
where $s \in [0,1]$ parametrizes the interpolation. In the limit 
$p \rightarrow \infty$, QAOA approaches a digitized version of 
adiabatic evolution under $H(s)$, provided the parameters 
$(\gamma_k,\beta_k)$ approximate the continuous schedule.

For shallow circuits, however, the evolution is necessarily 
nonadiabatic, and diabatic transitions can significantly reduce 
the overlap between the prepared state and the ground state of 
$H_C$.

QAOA performance has an evident dependence on the circuit depth and, as previously mentioned, it is observed that increasing depth improves the quality of the possible approximation (at the cost of increasing the parameter search space). Furthermore, finding appropriate solutions has a strong dependence on the ratio of a problem's constraints to variables (problem density). Hence, QAOA exhibits strong dependence on a problems density, and, for any fixed ansatz, there exists problem instances of high-density that appear not to be accessible. This feature persists as a fundamental limitation exhibited by QAOA~\cite{akshay_reachability_2020}.

Additionally, constrained optimization problems present several challenges for 
standard QAOA implementations. First, when constraints are implemented through penalty terms, 
the cost Hamiltonian contains energy scales associated with both 
the optimization objective and constraint enforcement \cite{bucher_efficient_2025}. Large 
penalty coefficients distort the spectrum of $H_C$ and can produce 
small effective gaps along the interpolation path, making it 
difficult for low-depth circuits to approximate the ground state.

Second, the standard transverse-field mixer does not preserve 
constraints such as fixed Hamming weight \cite{fuchs_constrained_2022}. As a result, the QAOA 
state may explore infeasible regions of the Hilbert space during 
the optimization. Although penalty terms suppress such states 
energetically, this mechanism can lead to inefficient exploration 
and slower convergence of the classical optimizer.

Finally, the dense interaction structure arising from covariance 
matrices in portfolio optimization generates a complex energy 
landscape with many competing configurations. In such settings, 
low-depth QAOA circuits may struggle to capture the correlations 
required to approximate the optimal portfolio \cite{he_alignment_2023}.

\subsection{Counterdiabatic Corrections}

From the perspective of adiabatic dynamics, the limitations above 
can be understood as consequences of diabatic transitions induced 
by the finite-depth evolution. When the interpolation between 
$H_M$ and $H_C$ encounters small spectral gaps, transitions out of 
the instantaneous ground state become likely, reducing the final 
state fidelity.

Counterdiabatic (CD) driving provides a principled mechanism for 
suppressing such transitions by augmenting the Hamiltonian with 
additional terms that cancel diabatic excitations \cite{guery-odelin_shortcuts_2019}. While the exact counterdiabatic Hamiltonian is generally nonlocal, variational approximations based on the adiabatic gauge potential offer a tractable route for incorporating these corrections into variational quantum circuits \cite{jarzynski_generating_2013}.



Consider a family of Hamiltonians $H(\lambda)$ parametrized by a 
control parameter $\lambda$. In adiabatic quantum evolution, the system 
is evolved according to $H(\lambda(t))$ with $\lambda$ slowly varying 
in time. Addition of a counterdiabatic Hamiltonian term to the adiabatic component results in

\begin{equation}
    H(\lambda) = (1 - \lambda(t))H_{i} + \lambda(t)H_{f} + f(\lambda)H_{CD},
\end{equation}
where $H_{i}$ and $H_{f}$ are the initial and problem Hamiltonians ($H_{AD} = (1 - \lambda(t))H_{i} + \lambda(t)H_{f}$), respectively, and $H_{CD}$ is the counterdiabatic term with scheduling $f(\lambda)$, which vanishes at the beginning and end of the protocol \cite{hegade_digitized_2022}.

Inclusion of the exact CD term allows following the instantaneous ground state of the original Hamiltonian at all times during the evolution. Therefore, the CD Hamiltonian term can be defined as,

\begin{equation}
    H_{CD} = \dot{\lambda}A_{\lambda},
\end{equation}
where $A_{\lambda}$ is known as the adiabatic gauge potential (AGP). When the evolution is very fast, $H_{CD}$ increases dramatically, and in the adiabatic limit this term vanishes. The adiabatic gauge potential is defined as

\begin{equation}
    A_{\lambda} = i\sum_{m}(1 - \ket{m(\lambda)}\bra{m(\lambda)})\ket{\partial m(\lambda)}\bra{m(\lambda)},
\end{equation}
corresponding to the nonadiabatic transitions between eigenstates $\ket{m(\lambda)}$ \cite{kolodrubetz_geometry_2017}. The AGP satisfies the condition 
\begin{equation}
    [i\partial H_{AD} - [A_{\lambda}, H_{AD}], H_{AD}] = 0.
\end{equation}

Obtaining the exact form for $A_{\lambda}$ is challenging since, in many cases, it contains exponentially many terms with nonlocal many-body interactions. In many cases, it suffices to employ the approximate form of the adiabatic gauge potential, which can be obtained from a nested commutator

\begin{equation}
        A_{\lambda}^{(l)} = i\sum_{k=1}^{l}\alpha_{k}(t)\underbrace{[H_{ad}, [H_{ad}, \dots [H_{ad},}_{2k-1} \partial_{\lambda}H_{ad}]]],
\end{equation}

where $\alpha_{k}(t)$ is the counterdiabatic coefficient. This coefficient is obtained by minimizing the operator distance between the exact gauge potential and the approximate gauge potential. In some sense, we construct the approximate gauge potential by restricting $A_{\lambda}$ to a variational operator ansatz

\begin{equation}
    A_{\lambda}^{*} = \sum_k \alpha_k(\lambda) O_k ,
\end{equation}

where $\{O_k\}$ is a chosen operator basis and $\alpha_k$ are 
variational coefficients.

Following the variational principle introduced in the context of 
adiabatic gauge potentials, the optimal coefficients are obtained by 
minimizing the Frobenius norm of the Hilbert-Schmidt operator

\begin{equation}
    G(A_{\lambda}) = 
    \partial_{\lambda} H_{AD} + i [A_{\lambda}, H_{AD}],
\end{equation}
which corresponds to minimizing the action
\begin{equation}
    \mathcal{S}(A_{\lambda}) =
    \mathrm{Tr}\left[ G(A_{\lambda})^2 \right].
\end{equation}

This yields a linear system of equations for the coefficients 
$\alpha_k$. The resulting operator $A_{\lambda}^{*}$ provides the 
best approximation to the exact gauge potential within the chosen 
operator subspace.

Once an approximate gauge potential has been obtained, to incorporate these corrections into a variational circuit, evolution is digitized in a manner analogous to QAOA. Specifically, the counterdiabatic QAOA ansatz augments the standard alternating structure with unitaries generated by the approximate gauge potential operators.

For circuit depth $p$, the resulting variational state is

\begin{equation}
\ket{\psi_{\mathrm{CD}}} =
\prod_{k=1}^{p}
\prod_{j} e^{-i \eta_{k,j} O_j}
e^{-i \beta_k H_M}
e^{-i \gamma_k H_C}
\ket{\psi_0},
\end{equation}
where $\eta_{k,j}$ are additional variational parameters associated 
with the counterdiabatic generators $O_j$.

These additional operators compensate for diabatic transitions that 
occur in shallow QAOA circuits, effectively increasing the expressive 
power of the variational ansatz without requiring a proportional 
increase in circuit depth.

A crucial ingredient in the above construction is the choice of 
operator basis $\{O_k\}$. In principle, the exact gauge potential 
can involve highly nonlocal Pauli strings, making a naive operator 
expansion impractical. However, the structure of the cost Hamiltonian 
$H_{C}$ strongly constrains the form of operators that 
appear in the nested commutator expansion of the gauge potential.

\subsection{Constrained Portfolio Optimization as a Quantum Hamiltonian}

Portfolio optimization is a canonical problem in quantitative finance, originating from the Markowitz mean-variance framework \cite{kerenidis_quantum_2019}. The goal is to construct a portfolio that balances expected return against risk, where risk is typically quantified by the covariance of asset returns.

Consider a universe of $N$ assets with expected return vector $\boldsymbol{\mu} = (\mu_1,\ldots,\mu_N)$ and covariance matrix $\Sigma$. In the classical mean-variance formulation, the objective is to minimize the risk-adjusted cost function

\begin{equation}
    C(\mathbf{x}) = \lambda \sum_{i,j=1}^{N} \Sigma_{ij} x_i x_j - \sum_{i=1}^{N} \mu_i x_i,
\end{equation}
where $x_i \in \{0,1\}$ denotes whether asset $i$ is included in the portfolio and $\lambda$ controls the tradeoff between return and risk. The first term penalizes correlated selections that increase portfolio variance, while the second term favors assets with high expected returns.

In many practical settings, the portfolio must satisfy a budget constraint specifying the number of selected assets. This can be expressed as

\begin{equation}
    \sum_{i=1}^{N} x_i = B ,
\end{equation}
where $B$ is the desired portfolio size.

One common approach for satisfying the budget is to enforce this constraint through a quadratic penalty term, yielding the unconstrained objective

\begin{equation}
    C_P(\mathbf{x}) =
    \lambda \sum_{i,j} \Sigma_{ij} x_i x_j
    - \sum_{i} \mu_i x_i
    + \alpha \left(\sum_i x_i - B\right)^2 ,
\end{equation}
where $\alpha$ is a penalty coefficient chosen large enough to ensure feasible solutions satisfy the constraint.

Alternatively, one may restrict the quantum evolution to the feasible 
subspace defined by $\sum_i x_i = B$, for example by designing mixers 
that preserve Hamming weight \cite{ruan_xy-mixer_2025}.

To encode the optimization problem in a quantum algorithm, we map the 
binary variables $x_i$ to spin-$1/2$ degrees of freedom using the 
standard transformation

\begin{equation}
    x_i = \frac{1 - Z_i}{2},
\end{equation}
where $Z_i$ is the Pauli-$Z$ operator acting on qubit $i$.

Substituting this relation into the cost function produces a Hamiltonian 
whose ground state corresponds to the optimal portfolio. Up to a 
constant energy shift, the resulting cost Hamiltonian takes the form

\begin{equation}
    H_C = 
    \sum_{i<j} J_{ij} Z_i Z_j
    + \sum_i h_i Z_i ,
\label{eq:portfolio_ising}
\end{equation}
where the coupling coefficients $J_{ij}$ arise from the covariance 
matrix and constraint penalties, and the local fields $h_i$ encode 
contributions from expected returns and linear terms generated by the 
constraint expansion \cite{rebentrost_quantum_2024, tang_comparative_2024}.

The Hamiltonian $H_C$ therefore has the structure of a fully connected 
Ising model with pairwise $ZZ$ interactions and local $Z$ fields. 
Explicitly, the interaction coefficients are determined by

\begin{equation}
    J_{ij} = 
    \frac{\lambda}{2} \Sigma_{ij},
\end{equation}
for $i < j$, up to convention and penalty terms, while the local fields depend on both the return vector and the constraint parameters. 

This form is directly compatible with QAOA and related variational 
algorithms, where the cost unitary is generated by $H_C$. Importantly, 
the dense interaction graph induced by the covariance matrix and the 
additional constraint penalties produces a complex energy landscape, 
making this class of problems a useful testbed for studying improved 
variational ansätze, such as counterdiabatic QAOA.

\section{Related Work}
Several well-known challenges limit the practical performance of QAOA, including the cost of repeated classical optimization, the rapid growth of the variational parameter space with circuit depth, and deteriorating scalability on large constrained instances \cite{abbas2024challenges}. A number of improvements have therefore been proposed to address these issues. Parameter transferability is one of the most effective acceleration mechanisms, as optimized angles can often be reused across related instances, depths, or problem sizes, reducing and sometimes largely avoiding expensive classical optimization \cite{galda2023similarity,falla2024graph,nguyen2025cross}. Other works exploit problem symmetries and invariant subspaces to identify redundant parameters and reduce the effective dimensionality of the QAOA ansatz \cite{tsvelikhovskiy2026reductions,tsvelikhovskiy2024equivariant,tsvelikhovskiy2024symmetries,shaydulin2021classical}. Initialization plays a similarly important role: warm-start strategies (e.g., problem-informed initial states, and structured parameter seeding) \cite{tate2023warm,egger2021warm}, and initialization schemes based on flexible distributions such as the beta-distribution-based \cite{kulshrestha2022beinit} can significantly improve optimization quality and trainability. To extend QAOA to larger instances, several hybrid methods combine it with decomposition frameworks that partition a large optimization problem into smaller subproblems and then coordinate their solutions \cite{maciejewski2024multilevel,bach2024mlqaoa}.

For optimization problems with hard constraints, the standard transverse-field mixer is generally insufficient because it explores infeasible regions of the Hilbert space. To address this limitation, Hadfield \emph{et al.} introduced the Quantum Alternating Operator Ansatz (QAOA$^+$), which generalizes QAOA by allowing problem-specific mixers that preserve feasibility \cite{hadfield_quantum_2019}. 
Among these, the $XY$ mixer has become one of the most important constructions for Hamming-weight constrained problems because it preserves particle number (or budget) exactly. 
Analytical and numerical studies of $XY$ mixers were further developed by Wang \emph{et al.}, who demonstrated their effectiveness for constrained optimization and one-hot encodings \cite{wang_x_2020}. 
Subsequent work has refined the theory and compilation of constrained mixers, including efficient Trotterized implementations and generalizations beyond the standard ring and complete-graph topologies \cite{fuchs_constrained_2022}.

These constrained mixer ideas have been particularly impactful in finance applications, especially portfolio optimization, where cardinality and budget constraints naturally define a fixed Hamming-weight subspace. 
Recent benchmarking studies have shown that $XY$-mixer QAOA provides significant advantages over unconstrained mixers for portfolio instances \cite{he_alignment_2023,brandhofer_benchmarking_2022}. 
In particular, the importance of initializing in the ground state of the $XY$ mixer (Dicke state) has been linked to improved trainability and better alignment with the adiabatic limit \cite{he_alignment_2023}.

Parallel to the development of QAOA, the quantum control and adiabatic computing communities have established the framework of \emph{counterdiabatic driving} (CD) and \emph{shortcuts to adiabaticity}, in which additional Hamiltonian terms suppress diabatic transitions during finite-time evolution \cite{guery-odelin_shortcuts_2019}. 
A central object in this theory is the adiabatic gauge potential (AGP), which generates the exact counterdiabatic correction. 
Because exact AGPs are typically intractable, Sels and Polkovnikov introduced a variational principle for constructing approximate local AGPs by minimizing an action based on the operator commutator structure \cite{sels_minimizing_2017}. 
This variational AGP framework has since become a standard route for practical counterdiabatic approximations in many-body systems.

The connection between QAOA and counterdiabatic driving has motivated several recent attempts to augment variational quantum optimization with AGP-inspired terms. 
These works generally show that incorporating approximate counterdiabatic corrections can improve optimization performance at low circuit depth by emulating faster adiabatic transport. 

\section{Methods}

\begin{figure*}[htbp]
\includegraphics[width=\textwidth, height=8cm]{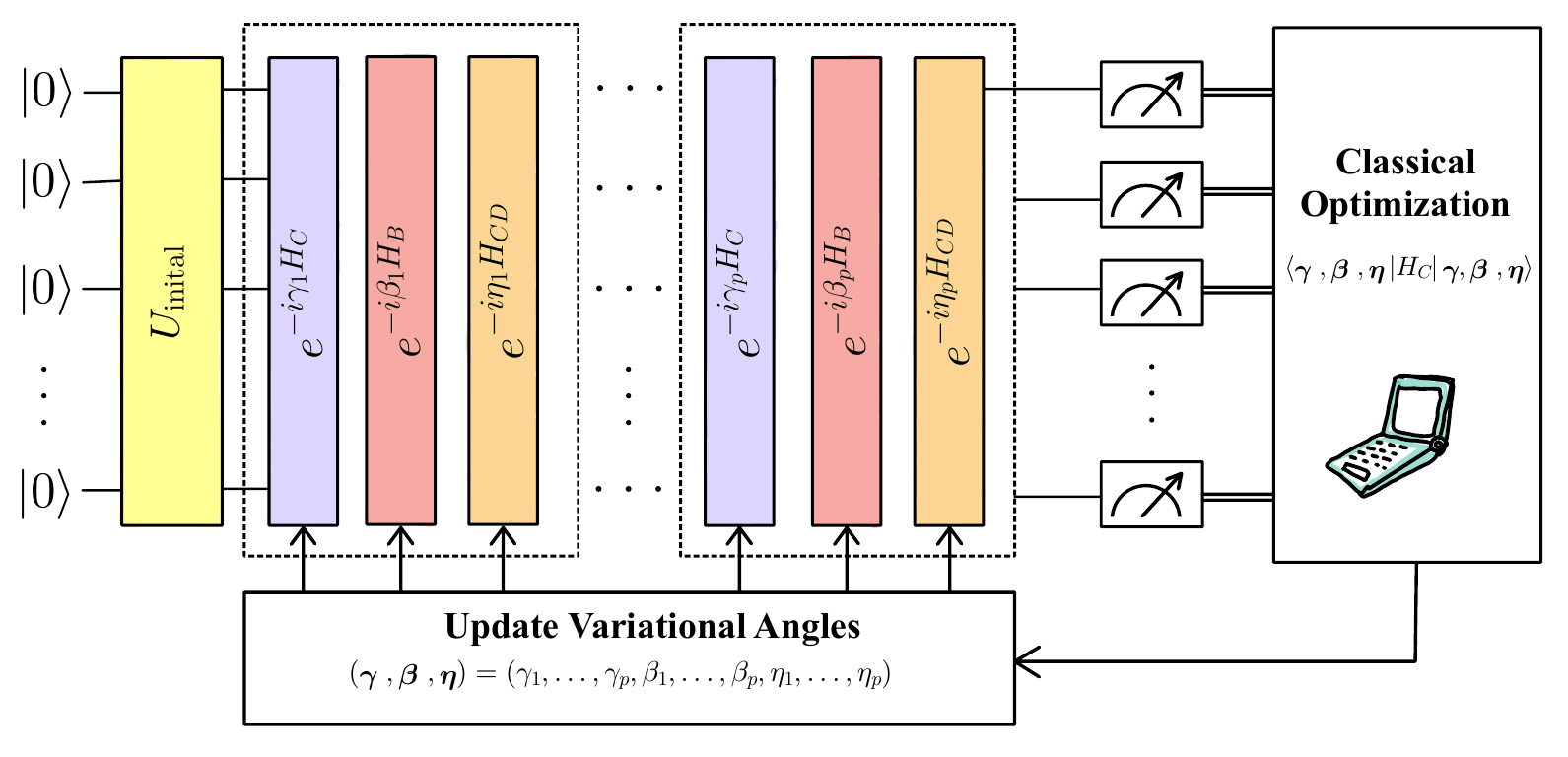}
\caption{\label{fig:schematic} Schematic of Counterdiabatic QAOA. For the portfolio optimization problem, the basis computational state is initialized to the fixed-Hamming-weight Dicke state. Subsequently, $p$ layers of the cost, mixer, and counterdiabatic unitaries are applied. A final measurement in the computational basis is made, and the variational parameters are optimized and updated classically.}
\end{figure*}

For the portfolio optimization problem, we wish to enforce the budget constraint $\sum_i x_i = B$ exactly. To this end, we restrict 
the dynamics to the fixed-Hamming-weight subspace using an XY mixer. 
The mixer Hamiltonian is defined as

\begin{equation}
    H_M^{XY} = \sum_{i<j} \left( X_i X_j + Y_i Y_j \right),
\end{equation}
which preserves the total excitation number. As a result, the evolution generated by $H_M^{XY}$ remains entirely within the feasible subspace of valid portfolios.

The QAOA circuit is initialized in a Dicke state of Hamming weight $B$,

\begin{equation}
    \ket{D^N_B} = \binom{N}{B}^{-1/2}
    \sum_{\substack{z \in \{0,1\}^N \\ |z| = B}} \ket{z},
\end{equation}
which is an equal superposition over all feasible configurations. This state can be prepared efficiently using known state preparation 
protocols and serves as natural eigenstate of $H_M^{XY}$ within the 
fixed-weight subspace \cite{li_preparation_2025}.

We consider the interpolating Hamiltonian

\begin{equation}
    H(\lambda) = (1-\lambda) H_M^{XY} + \lambda H_C,
\end{equation}
where, for the portfolio optimization problem, $H_{C}$ has the form of Eq. \ref{eq:portfolio_ising}. The commutator

\begin{equation}
    [H_C, H_M^{XY}]
\end{equation}
generates operators of the form
\begin{equation}
    X_i Y_j Z_k, \quad Y_i X_j Z_k,
\end{equation}
which correspond to three-body Pauli strings. Motivated by this structure, we truncate the gauge potential to the form

\begin{equation}
    A_\lambda^{*}
    =
    \sum_{i<j} \alpha_{ij} \left( X_i Y_j - Y_i X_j \right)
    +
    \sum_{i<j<k} \beta_{ijk} \, O_{ijk},
\end{equation}
where $O_{ijk}$ denotes three-body Pauli strings of the form

\begin{equation}
    O_{ijk} \in \{ X_i Y_j Z_k - Y_i X_j Z_k, \ \text{permutations} \}.
\end{equation}

The first term corresponds to two-body generators already present in 
the XY mixer algebra, while the second captures leading-order 
corrections arising from interactions with the cost Hamiltonian.

To illustrate how higher-body operators arise in the gauge potential, 
we consider a representative commutator between a term in the cost 
Hamiltonian and a term in the XY mixer.

Let $H_C \supset J_{ij} Z_i Z_j$ and 
$H_M^{XY} \supset X_i X_k + Y_i Y_k$, where $i,j,k$ are distinct. 
Using Pauli commutation relations, we compute

\begin{align}
    [Z_i Z_j, X_i X_k]
    &= Z_j [Z_i, X_i] X_k \\
    &= Z_j (2i Y_i) X_k \\
    &= 2i \, Y_i X_k Z_j.
\end{align}

Similarly,

\begin{align}
    [Z_i Z_j, Y_i Y_k]
    &= Z_j [Z_i, Y_i] Y_k \\
    &= Z_j (-2i X_i) Y_k \\
    &= -2i \, X_i Y_k Z_j.
\end{align}

Combining these contributions, we obtain

\begin{equation}
    [Z_i Z_j, X_i X_k + Y_i Y_k]
    =
    2i \left( Y_i X_k - X_i Y_k \right) Z_j,
\end{equation}
which is a three-body Pauli operator. This example demonstrates that commutators between the cost Hamiltonian 
and the XY mixer naturally generate operators of the form 
$X_i Y_k Z_j - Y_i X_k Z_j$, motivating the inclusion of three-body 
terms in the gauge potential ansatz.

The coefficients $\alpha_{ij}$ and $\beta_{ijk}$ are obtained by 
minimizing the action

\begin{equation}
    \mathcal{S} = \mathrm{Tr}\left[
    \left( \partial_\lambda H + i [A_\lambda, H] \right)^2
    \right].
\end{equation}

Substituting the ansatz for $A_\lambda^{*}$ into this expression yields 
a system of linear equations of the form

\begin{equation}
    \sum_{l} M_{kl} \, c_l = v_k,
\end{equation}
where $c_l \in \{\alpha_{ij}, \beta_{ijk}\}$, and
\begin{align}
    M_{kl} &= -\mathrm{Tr}\left( [O_k, H][O_l, H] \right), \\
    v_k &= i \, \mathrm{Tr}\left( O_k [H, \partial_\lambda H] \right).
\end{align}

This linear system determines the optimal coefficients within the 
chosen operator subspace.

For the portfolio Hamiltonian considered here, the structure of $H_C$ 
implies that the coefficients $\alpha_{ij}$ depend primarily on local 
fields and pairwise couplings involving qubits $i$ and $j$, while the 
three-body coefficients $\beta_{ijk}$ encode higher-order correlations 
induced by the covariance matrix.

To implement this evolution in a variational circuit, we digitize the 
time evolution into $p$ layers, yielding the ansatz

\begin{equation}
\ket{\psi_{\mathrm{CD}}} =
\prod_{k=1}^{p}
e^{-i \eta_{k}H_{CD}}
e^{-i \beta_k H_M^{XY}}
e^{-i \gamma_k H_C}
\ket{D^N_B},
\end{equation}
where standard QAOA is augmented by three-body counterdiabatic generators. While the non-counterdiabatic XY-mixer approach preserves Hamming weight, adding these counterdiabatic contributions will result in leakage into the infeasible subspace. This is further analyzed in the following section.

The additional parameters $\eta_{k}$ allow the circuit to approximate 
counterdiabatic evolution while retaining the flexibility of a 
variational ansatz. In practice, we find that truncating the operator 
expansion to include only up to three-body terms provides a favorable 
tradeoff between expressibility and circuit complexity.

\section{Numerical Results}

We now benchmark the performance of counterdiabatic QAOA for constrained portfolio optimization and compare it against standard QAOA implemented with the same XY mixer and fixed-Hamming-weight initialization (Dicke state).

Unless otherwise stated, all numerical simulations are performed on randomly generated portfolio instances, using Qiskit Finance, with $N=12$ assets, expected return vector $\boldsymbol{\mu}$ sampled from a uniform distribution, and covariance matrices $\Sigma$ generated as dense positive semidefinite matrices. The portfolio size is fixed to Hamming weight $B=4$, and all simulations are restricted to this feasible subspace.

Furthermore, we employ the conditional value-at-risk (CVAR) technique, which deviates from the traditional sampling of the mean in measurement outcomes. We use a CVAR approach since problems with a diagonal Hamiltonian, such as combinatorial optimization problems, the sample mean may be a poor choice because when $H$ is diagonal, there exists a ground state which is a basis state \cite{barkoutsos_improving_2020}.

All numerical simulations were performed by augmenting \textit{OpenQuantumComputing}'s implementation of QAOA \cite{fuchs2024qaoa} with a counterdiabatic routine.

We first study the approximation ratio as a function of QAOA depth $p$. For each instance, we define

\begin{equation}
    r = \frac{\langle H_C \rangle - E_{\max}}
    {E_{\min} - E_{\max}},
\end{equation}
where $E_{\min}$ and $E_{\max}$ denote the minimum and maximum energies within the feasible subspace.

\begin{figure}[htbp]
\includegraphics[width=0.5\textwidth]{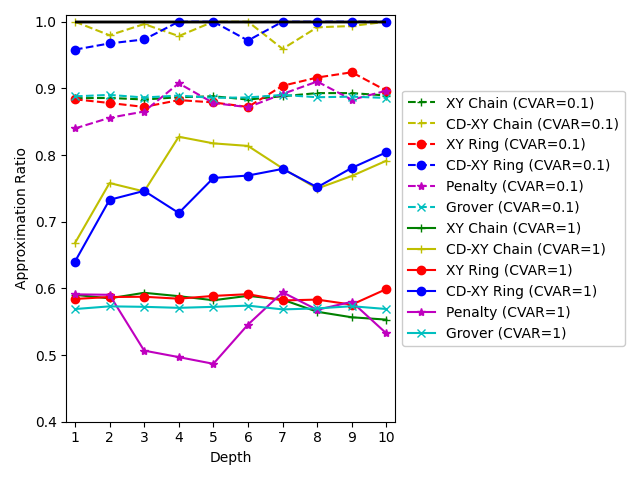}
\caption{\label{fig:approx_ratio_vs_depth} Comparison of approximation ratios for XY mixers (with and without CD contributions), penalty-based constrained QAOA, and Grover mixer QAOA for different conditional values at risk.}
\end{figure}

Figure~\ref{fig:approx_ratio_vs_depth} shows a comparison in approximation ratio for different approaches to constrained portfolio optimization. The methods presented are standard XY-mixer QAOA, XY-mixer CD QAOA (our approach), Grover mixer QAOA \cite{bartschi_grover_2020}, and penalty-based QAOA. For the XY mixers, we employ both chain and ring structures \cite{wang_x_2020}. Furthermore, this figure shows approximation ratios for different conditional values at risk. The figure shows that CD-QAOA consistently achieves higher approximation ratios at fixed $p$. For all methods, a smaller CVAR value results in a better approximation ratio.

To quantify the ability of the variational circuits to prepare the optimal portfolio state, we compute the success probability

\begin{equation}
    P_{\mathrm{GS}} = |\langle \psi_{\mathrm{opt}} | \psi \rangle|^2,
\end{equation}
where $\ket{\psi_{\mathrm{opt}}}$ is the exact ground state of the portfolio Hamiltonian in the feasible subspace.

\begin{figure}[htbp]
\includegraphics[width=0.5\textwidth]{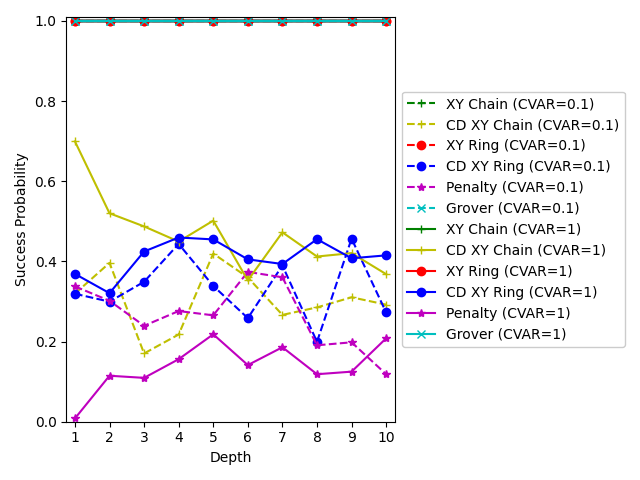}
\caption{\label{fig:success_prob} Comparison of success probabilities for XY mixers (with and without CD contributions), penalty-based constrained QAOA, and Grover mixer QAOA for different conditional values at risk.}
\end{figure}

A key aspect of Fig.~\ref{fig:success_prob} is that it points to a limitation of including counterdiabatic driving to XY-mixer QAOA. By constructing the CD Hamiltonian with three-body terms, we are effectively leaking into the infeasible subspace because these terms are not Hamming-weight preserving. By comparison, an XY-mixer approach that does not include a CD term always results in a success probability of $1$, as all sampled bit strings are feasible solutions. Nevertheless, we see that, in general, our approach produces better success probabilities compared to a penalty-based approach, which is also Hamming-weight non-preserving.

Including three-body CD correction terms to our XY-mixer architecture increases the complexity of the ansatz. This is evidenced in an increased optimization runtime as well as an increase in the transpiled circuit depth and number of CNOT gates. 

\begin{figure}[htbp]
\includegraphics[width=0.45\textwidth]{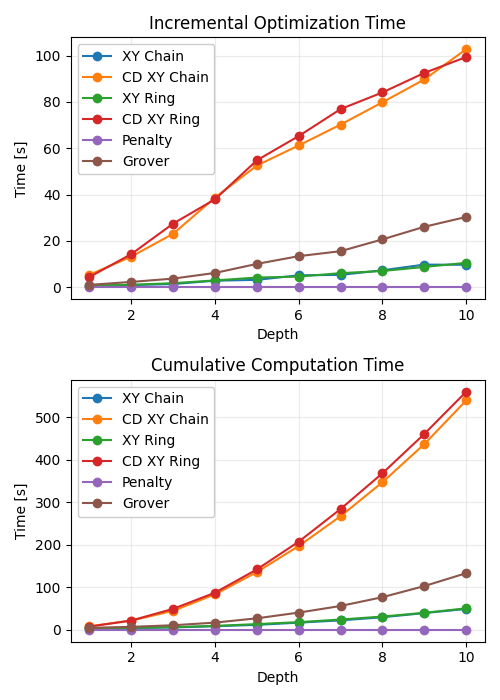}
\caption{\label{fig:runtime} Incremental and cumulative runtime for all constrained optimization methods at a fixed CVAR value of 1.}
\end{figure}

\begin{figure}[htbp]
\includegraphics[width=0.45\textwidth]{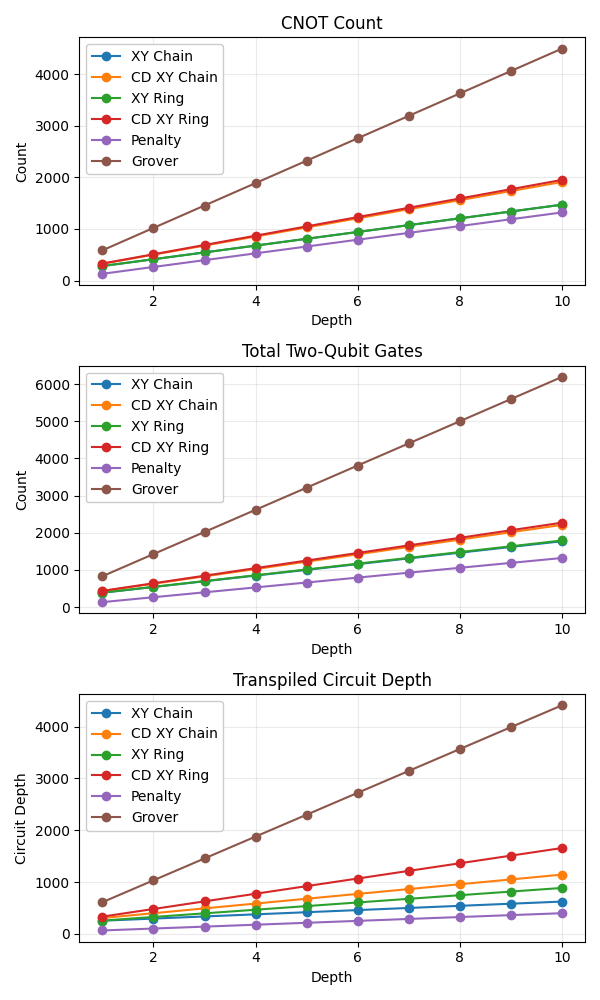}
\caption{\label{fig:metrics} Scaling of CNOT gate count, total two-qubit gate count, and transpiled circuit depth as a function of QAOA depth for all constrained optimization methods at a fixed CVAR value of 1.}
\end{figure}

Figures~\ref{fig:runtime} and \ref{fig:metrics} illustrate the scaling of runtime, CNOT gate count, total two-qubit gate count, and transpiled circuit depth for increasing QAOA depth. As expected, the runtime in the counterdiabatic regime exceeds runtime for any other method; not only is there a quantum overhead in the number of necessary gates, but also an additional variational parameter to be optimized, leading to classical overhead as well. This scalability issue could be addressed by, for example, warm starting the QAOA instance \cite{kordonowy_lie_2025, bucher_constrained_2026}.

On the other hand, while we see an increase in the number of CNOT and total two-qubit gates, and in the transpiled circuit depth, the CD approach remains scalable when compared to Grover-mixer QAOA. A possible solution to reduce the total number of two-qubit gates in the CD regime is to adaptively select a subset of the three-body CD operator pool, akin to ADAPT-QAOA \cite{zhu_adaptive_2022}.

\section{Discussion}

The numerical results presented above demonstrate that incorporating 
counterdiabatic corrections into constrained QAOA provides systematic 
performance improvements across portfolio instances, particularly in 
the shallow-circuit regime. From a dynamical perspective, this 
improvement can be understood as a direct consequence of suppressing 
diabatic transitions along the interpolation path between the XY mixer 
and the portfolio cost Hamiltonian.

The use of an XY mixer restricts the dynamics to the fixed-Hamming-weight 
subspace, ensuring exact preservation of the budget constraint. And while adding a CD term causes leakage to the infeasible subspace, the success probability is improved over a penalty-based method. Within 
this constrained manifold, the interpolation encounters nontrivial 
spectral structure generated by the dense covariance matrix and the 
risk-return tradeoff terms. In particular, near-degenerate excited 
states frequently arise from competing portfolio configurations with 
similar variance profiles. At low circuit depth, standard QAOA lacks 
sufficient expressive power to resolve these avoided crossings, leading 
to diabatic population transfer into suboptimal feasible states.

The variationally constructed adiabatic gauge potential compensates for 
these transitions by introducing operators that locally rotate the 
instantaneous eigenbasis of the interpolating Hamiltonian. The 
commutator-based derivation makes this mechanism explicit: the leading 
corrections generated by the XY mixer and the pairwise Ising cost terms 
naturally produce three-body Pauli strings. These terms encode 
correlated excitation transport conditioned on the state of neighboring 
qubits, allowing the circuit to capture higher-order dependencies in 
the covariance graph.

This observation provides important physical intuition for why the 
three-body truncation performs particularly well. While the original 
portfolio Hamiltonian contains only pairwise $ZZ$ couplings, the 
instantaneous eigenbasis rotates in a correlated many-body manner under 
the XY dynamics. The three-body counterdiabatic terms represent the 
lowest-order operators capable of capturing this correlated basis 
rotation, explaining their disproportionate impact on shallow-depth 
performance.

More broadly, our results suggest that constrained combinatorial 
optimization problems possess a natural operator hierarchy. Rather 
than relying on heuristic ansatz design, the counterdiabatic framework 
provides a systematic route to identifying the most physically relevant 
generators needed to suppress diabatic errors. This perspective may be 
particularly valuable for other fixed-subspace optimization problems, 
including scheduling, routing, and cardinality-constrained quadratic 
programs.

An important practical implication concerns circuit complexity. Although 
the full gauge potential generally contains operators of increasing 
Pauli weight, our results indicate that truncation at three-body terms 
already captures the dominant corrections for the portfolio instances 
studied here. This suggests a favorable expressibility-to-complexity 
tradeoff in which significant gains can be achieved without the 
prohibitive overhead associated with deeper standard QAOA circuits.

Finally, the observed improvements in classical optimizer convergence 
suggest that counterdiabatic corrections reshape the variational 
landscape in addition to improving circuit expressibility. A more 
detailed analysis of trainability, gradient concentration, and spectral 
geometry in constrained CD-QAOA remains an important direction for 
future work.

\section{Conclusion}

In this work, we introduced a counterdiabatic extension of the Quantum 
Approximate Optimization Algorithm for constrained portfolio 
optimization. By combining an XY mixer, Dicke-state initialization, and 
a variational approximation to the adiabatic gauge potential, we 
constructed an ansatz that systematically improves 
upon standard QAOA.

A key result of this work is the explicit construction of 
the gauge-potential operator pool generated by the portfolio cost 
Hamiltonian and the XY mixer. This analysis shows that the leading 
counterdiabatic corrections naturally include both two-body and 
three-body Pauli operators, with the latter playing a particularly 
important role in capturing correlated diabatic effects within the 
feasible subspace. Our numerical simulations demonstrate that these counterdiabatic terms 
lead to improved approximation ratios  and superior shallow-depth 
scaling across portfolio instances with dense covariance structure. 

Beyond the specific application to portfolio optimization, our results 
establish a general framework for enhancing constrained variational 
quantum algorithms through systematically derived counterdiabatic 
corrections. The connection between adiabatic gauge potentials and structured optimization Hamiltonians provides a broadly applicable methodology for constructing physically motivated ansätze in constrained Hilbert spaces.

Future work includes extending this framework to larger financial 
optimization models using classically driven decomposition \cite{ushijima2021multilevel},  operator truncation strategies with adaptive \cite{zhu2022adaptive} and generative learning \cite{tyagin2025qaoa} methods, and 
hardware-efficient decompositions of the three-body counterdiabatic 
terms. More generally, the present results suggest that combining 
constraint-preserving mixers with truncated gauge-potential expansions 
may offer a scalable path toward improved quantum optimization on 
near-term devices.

\section*{Acknowledgment}
This work was supported in part by NSF award \#2427042.

\bibliographystyle{unsrt}
\bibliography{ccd_qaoa_article_qce,ilya-biblio}

@misc{farhi_quantum_2014,
	title = {A {Quantum} {Approximate} {Optimization} {Algorithm}},
	url = {http://arxiv.org/abs/1411.4028},
	language = {en},
	urldate = {2023-04-27},
	publisher = {arXiv},
	author = {Farhi, Edward and Goldstone, Jeffrey and Gutmann, Sam},
	month = nov,
	year = {2014},
	note = {arXiv:1411.4028 [quant-ph]},
}

@misc{tang_comparative_2024,
	title = {Comparative analysis of diverse methodologies for portfolio optimization leveraging quantum annealing techniques},
	url = {http://arxiv.org/abs/2403.02599},
	language = {en},
	urldate = {2024-08-21},
	publisher = {arXiv},
	author = {Tang, Zhijie and Dou, Alex Lu and Bishwas, Arit Kumar},
	month = jul,
	year = {2024},
	note = {arXiv:2403.02599 [quant-ph]},
}

@inproceedings{kerenidis_quantum_2019,
	address = {Zurich Switzerland},
	title = {Quantum {Algorithms} for {Portfolio} {Optimization}},
	isbn = {978-1-4503-6732-5},
	url = {https://dl.acm.org/doi/10.1145/3318041.3355465},
	doi = {10.1145/3318041.3355465},
	language = {en},
	urldate = {2024-08-21},
	booktitle = {Proceedings of the 1st {ACM} {Conference} on {Advances} in {Financial} {Technologies}},
	publisher = {ACM},
	author = {Kerenidis, Iordanis and Prakash, Anupam and Szilágyi, Dániel},
	month = oct,
	year = {2019},
	pages = {147--155},
}

@article{brandhofer_benchmarking_2022,
	title = {Benchmarking the performance of portfolio optimization with {QAOA}},
	volume = {22},
	issn = {1573-1332},
	url = {https://link.springer.com/10.1007/s11128-022-03766-5},
	doi = {10.1007/s11128-022-03766-5},
	language = {en},
	number = {1},
	urldate = {2024-08-21},
	journal = {Quantum Inf Process},
	author = {Brandhofer, Sebastian and Braun, Daniel and Dehn, Vanessa and Hellstern, Gerhard and Hüls, Matthias and Ji, Yanjun and Polian, Ilia and Bhatia, Amandeep Singh and Wellens, Thomas},
	month = dec,
	year = {2022},
	pages = {25},
}

@article{rebentrost_quantum_2024,
	title = {Quantum {Computational} {Finance}: {Quantum} {Algorithm} for {Portfolio} {Optimization}},
	issn = {0933-1875, 1610-1987},
	shorttitle = {Quantum {Computational} {Finance}},
	url = {https://link.springer.com/10.1007/s13218-024-00870-9},
	doi = {10.1007/s13218-024-00870-9},
	language = {en},
	urldate = {2024-08-22},
	journal = {Künstl Intell},
	author = {Rebentrost, Patrick and Lloyd, Seth},
	month = aug,
	year = {2024},
}

@article{fuchs_constrained_2022,
	title = {Constrained mixers for the quantum approximate optimization algorithm},
	volume = {15},
	issn = {1999-4893},
	url = {http://arxiv.org/abs/2203.06095},
	doi = {10.3390/a15060202},
	language = {en},
	number = {6},
	urldate = {2025-06-03},
	journal = {Algorithms},
	author = {Fuchs, Franz G. and Lye, Kjetil Olsen and Nilsen, Halvor Møll and Stasik, Alexander J. and Sartor, Giorgio},
	month = jun,
	year = {2022},
	note = {arXiv:2203.06095 [quant-ph]},
	pages = {202},
}

@article{amin_consistency_2009,
	title = {Consistency of the {Adiabatic} {Theorem}},
	volume = {102},
	copyright = {http://link.aps.org/licenses/aps-default-license},
	issn = {0031-9007, 1079-7114},
	url = {https://link.aps.org/doi/10.1103/PhysRevLett.102.220401},
	doi = {10.1103/PhysRevLett.102.220401},
	language = {en},
	number = {22},
	urldate = {2025-06-03},
	journal = {Phys. Rev. Lett.},
	author = {Amin, M. H. S.},
	month = jun,
	year = {2009},
	pages = {220401},
}

@inproceedings{bartschi_grover_2020,
	address = {Denver, CO, USA},
	title = {Grover {Mixers} for {QAOA}: {Shifting} {Complexity} from {Mixer} {Design} to {State} {Preparation}},
	copyright = {https://ieeexplore.ieee.org/Xplorehelp/downloads/license-information/USG.html},
	isbn = {978-1-7281-8969-7},
	shorttitle = {Grover {Mixers} for {QAOA}},
	url = {https://ieeexplore.ieee.org/document/9259965/},
	doi = {10.1109/QCE49297.2020.00020},
	language = {en},
	urldate = {2025-06-03},
	booktitle = {2020 {IEEE} {International} {Conference} on {Quantum} {Computing} and {Engineering} ({QCE})},
	publisher = {IEEE},
	author = {Bartschi, Andreas and Eidenbenz, Stephan},
	month = oct,
	year = {2020},
	pages = {72--82},
}

@article{hegade_digitized_2022,
	title = {Digitized counterdiabatic quantum optimization},
	volume = {4},
	issn = {2643-1564},
	url = {https://link.aps.org/doi/10.1103/PhysRevResearch.4.L042030},
	doi = {10.1103/PhysRevResearch.4.L042030},
	language = {en},
	number = {4},
	urldate = {2025-06-03},
	journal = {Phys. Rev. Research},
	author = {Hegade, Narendra N. and Chen, Xi and Solano, Enrique},
	month = nov,
	year = {2022},
	pages = {L042030},
}

@article{guery-odelin_shortcuts_2019,
	title = {Shortcuts to adiabaticity: {Concepts}, methods, and applications},
	volume = {91},
	issn = {0034-6861, 1539-0756},
	shorttitle = {Shortcuts to adiabaticity},
	url = {https://link.aps.org/doi/10.1103/RevModPhys.91.045001},
	doi = {10.1103/RevModPhys.91.045001},
	language = {en},
	number = {4},
	urldate = {2025-06-03},
	journal = {Rev. Mod. Phys.},
	author = {Guéry-Odelin, D. and Ruschhaupt, A. and Kiely, A. and Torrontegui, E. and Martínez-Garaot, S. and Muga, J.G.},
	month = oct,
	year = {2019},
	pages = {045001},
}

@article{sels_minimizing_2017,
	title = {Minimizing irreversible losses in quantum systems by local counterdiabatic driving},
	volume = {114},
	issn = {0027-8424, 1091-6490},
	url = {https://pnas.org/doi/full/10.1073/pnas.1619826114},
	doi = {10.1073/pnas.1619826114},
	language = {en},
	number = {20},
	urldate = {2025-06-03},
	journal = {Proc. Natl. Acad. Sci. U.S.A.},
	author = {Sels, Dries and Polkovnikov, Anatoli},
	month = may,
	year = {2017},
}

@article{hadfield_quantum_2019,
	title = {From the {Quantum} {Approximate} {Optimization} {Algorithm} to a {Quantum} {Alternating} {Operator} {Ansatz}},
	volume = {12},
	issn = {1999-4893},
	url = {http://arxiv.org/abs/1709.03489},
	doi = {10.3390/a12020034},
	language = {en},
	number = {2},
	urldate = {2025-06-15},
	journal = {Algorithms},
	author = {Hadfield, Stuart and Wang, Zhihui and O'Gorman, Bryan and Rieffel, Eleanor G. and Venturelli, Davide and Biswas, Rupak},
	month = feb,
	year = {2019},
	note = {arXiv:1709.03489 [quant-ph]},
	pages = {34},
}

@misc{bucher_efficient_2025,
	title = {Efficient {QAOA} {Architecture} for {Solving} {Multi}-{Constrained} {Optimization} {Problems}},
	url = {http://arxiv.org/abs/2506.03115},
	doi = {10.48550/arXiv.2506.03115},
	language = {en},
	urldate = {2025-06-15},
	publisher = {arXiv},
	author = {Bucher, David and Porawski, Daniel and Janetschek, Maximilian and Stein, Jonas and O'Meara, Corey and Cortiana, Giorgio and Linnhoff-Popien, Claudia},
	month = jun,
	year = {2025},
	note = {arXiv:2506.03115 [quant-ph]},
}

@article{jarzynski_generating_2013,
	title = {Generating shortcuts to adiabaticity in quantum and classical dynamics},
	volume = {88},
	copyright = {http://link.aps.org/licenses/aps-default-license},
	issn = {1050-2947, 1094-1622},
	url = {https://link.aps.org/doi/10.1103/PhysRevA.88.040101},
	doi = {10.1103/PhysRevA.88.040101},
	language = {en},
	number = {4},
	urldate = {2025-06-27},
	journal = {Phys. Rev. A},
	author = {Jarzynski, Christopher},
	month = oct,
	year = {2013},
	pages = {040101},
}

@article{cepaite_counterdiabatic_2023,
	title = {Counterdiabatic {Optimized} {Local} {Driving}},
	volume = {4},
	issn = {2691-3399},
	url = {https://link.aps.org/doi/10.1103/PRXQuantum.4.010312},
	doi = {10.1103/PRXQuantum.4.010312},
	language = {en},
	number = {1},
	urldate = {2025-06-27},
	journal = {PRX Quantum},
	author = {Čepaitė, Ieva and Polkovnikov, Anatoli and Daley, Andrew J. and Duncan, Callum W.},
	month = jan,
	year = {2023},
	pages = {010312},
}

@misc{li_preparation_2025,
	title = {Preparation of {Hamming}-{Weight}-{Preserving} {Quantum} {States} with {Log}-{Depth} {Quantum} {Circuits}},
	url = {http://arxiv.org/abs/2508.14470},
	doi = {10.48550/arXiv.2508.14470},
	language = {en},
	urldate = {2025-09-12},
	publisher = {arXiv},
	author = {Li, Yu and Tian, Guojing and He, Xiaoyu and Sun, Xiaoming},
	month = aug,
	year = {2025},
	note = {arXiv:2508.14470 [quant-ph]},
}

@misc{kordonowy_lie_2025,
	title = {The {Lie} {Algebra} of {XY}-mixer {Topologies} and {Warm} {Starting} {QAOA} for {Constrained} {Optimization}},
	url = {http://arxiv.org/abs/2505.18396},
	doi = {10.48550/arXiv.2505.18396},
	language = {en},
	urldate = {2025-10-30},
	publisher = {arXiv},
	author = {Kordonowy, Steven and Leipold, Hannes},
	month = sep,
	year = {2025},
	note = {arXiv:2505.18396 [quant-ph]},
}

@article{barkoutsos_improving_2020,
	title = {Improving {Variational} {Quantum} {Optimization} using {CVaR}},
	volume = {4},
	issn = {2521-327X},
	url = {http://arxiv.org/abs/1907.04769},
	doi = {10.22331/q-2020-04-20-256},
	language = {en},
	urldate = {2025-10-30},
	journal = {Quantum},
	author = {Barkoutsos, Panagiotis Kl and Nannicini, Giacomo and Robert, Anton and Tavernelli, Ivano and Woerner, Stefan},
	month = apr,
	year = {2020},
	note = {arXiv:1907.04769 [quant-ph]},
	pages = {256},
}

@article{wang_x_2020,
	title = {X {Y} mixers: {Analytical} and numerical results for the quantum alternating operator ansatz},
	volume = {101},
	issn = {2469-9926, 2469-9934},
	shorttitle = {X {Y} mixers},
	url = {https://link.aps.org/doi/10.1103/PhysRevA.101.012320},
	doi = {10.1103/PhysRevA.101.012320},
	language = {en},
	number = {1},
	urldate = {2026-04-07},
	journal = {Phys. Rev. A},
	author = {Wang, Zhihui and Rubin, Nicholas C. and Dominy, Jason M. and Rieffel, Eleanor G.},
	month = jan,
	year = {2020},
	pages = {012320},
}

@article{akshay_reachability_2020,
	title = {Reachability {Deficits} in {Quantum} {Approximate} {Optimization}},
	volume = {124},
	issn = {0031-9007, 1079-7114},
	url = {https://link.aps.org/doi/10.1103/PhysRevLett.124.090504},
	doi = {10.1103/PhysRevLett.124.090504},
	language = {en},
	number = {9},
	urldate = {2026-04-07},
	journal = {Phys. Rev. Lett.},
	author = {Akshay, V. and Philathong, H. and Morales, M.E.S. and Biamonte, J.D.},
	month = mar,
	year = {2020},
	pages = {090504},
}

@article{ruan_xy-mixer_2025,
	title = {{XY}-mixer ansatz assisted by counterdiabatic driving for combinational optimization},
	volume = {7},
	issn = {2643-1564},
	url = {https://link.aps.org/doi/10.1103/PhysRevResearch.7.013243},
	doi = {10.1103/PhysRevResearch.7.013243},
	language = {en},
	number = {1},
	urldate = {2026-04-07},
	journal = {Phys. Rev. Research},
	author = {Ruan, Yue and Chen, Pengyue and Li, Qi and Yang, Ling and Yuan, Zhiqiang and Xue, Xiling and Li, Xi and Liu, Zhihao},
	month = mar,
	year = {2025},
	pages = {013243},
}

@article{he_alignment_2023,
	title = {Alignment between initial state and mixer improves {QAOA} performance for constrained optimization},
	volume = {9},
	issn = {2056-6387},
	url = {https://www.nature.com/articles/s41534-023-00787-5},
	doi = {10.1038/s41534-023-00787-5},
	language = {en},
	number = {1},
	urldate = {2026-04-07},
	journal = {npj Quantum Inf},
	author = {He, Zichang and Shaydulin, Ruslan and Chakrabarti, Shouvanik and Herman, Dylan and Li, Changhao and Sun, Yue and Pistoia, Marco},
	month = nov,
	year = {2023},
	pages = {121},
}

@article{saleem_approaches_2023,
	title = {Approaches to {Constrained} {Quantum} {Approximate} {Optimization}},
	volume = {4},
	issn = {2661-8907},
	url = {https://link.springer.com/10.1007/s42979-022-01638-4},
	doi = {10.1007/s42979-022-01638-4},
	language = {en},
	number = {2},
	urldate = {2026-04-07},
	journal = {SN COMPUT. SCI.},
	author = {Saleem, Zain H. and Tomesh, Teague and Tariq, Bilal and Suchara, Martin},
	month = jan,
	year = {2023},
	pages = {183},
}

@misc{bucher_constrained_2026,
	title = {Constrained {Quantum} {Optimization} via {Iterative} {Warm}-{Start} {XY}-{Mixers}},
	url = {http://arxiv.org/abs/2604.02083},
	doi = {10.48550/arXiv.2604.02083},
	language = {en},
	urldate = {2026-04-16},
	publisher = {arXiv},
	author = {Bucher, David and Janetschek, Maximilian and Poppel, Michael and Stein, Jonas and Linnhoff-Popien, Claudia and Feld, Sebastian},
	month = apr,
	year = {2026},
	note = {arXiv:2604.02083 [quant-ph]},
}

@article{zhu_adaptive_2022,
	title = {Adaptive quantum approximate optimization algorithm for solving combinatorial problems on a quantum computer},
	volume = {4},
	issn = {2643-1564},
	url = {https://link.aps.org/doi/10.1103/PhysRevResearch.4.033029},
	doi = {10.1103/PhysRevResearch.4.033029},
	language = {en},
	number = {3},
	urldate = {2026-04-16},
	journal = {Phys. Rev. Research},
	author = {Zhu, Linghua and Tang, Ho Lun and Barron, George S. and Calderon-Vargas, F. A. and Mayhall, Nicholas J. and Barnes, Edwin and Economou, Sophia E.},
	month = jul,
	year = {2022},
	pages = {033029},
}

@article{ZWCHL,
   author={Zhou, L and  Wang, S. and  Choi, S. and  Hannes, P. and  Lukin, M.},
   title={Quantum approximate optimization algorithm: Performance, mechanism, and implementation on near-term devices},
   journal={Phys. Rev. X},
   volume={10},
   year={2020}
}

@article{LTWS,
   author={Larocca, M. 
   and  Thanasilp, S. 
   and  Wang, S. 
   and  Sharma, K. 
   and  Biamonte, J. 
   and  Coles, P.~J. 
   and  Cincio, L.
   and  McClean, J.~R.
   and  Holmes, Z.},
   title={Barren plateaus in variational quantum computing},
   journal={Nat. Rev. Phys.},
   pages={1--16},
   year={2025}
}

@software{fuchs2024qaoa,
  author       = {Franz Georg Fuchs},
  title        = {{QAOA}: A Modular Python Library for the Quantum Approximate Optimization Algorithm},
  year         = {2024},
  url          = {https://github.com/OpenQuantumComputing/QAOA},
  note         = {Version 2.0.0}
}

@article{kolodrubetz_geometry_2017,
	title = {Geometry and non-adiabatic response in quantum and classical systems},
	volume = {697},
	issn = {03701573},
	url = {https://linkinghub.elsevier.com/retrieve/pii/S0370157317301989},
	doi = {10.1016/j.physrep.2017.07.001},
	language = {en},
	urldate = {2025-06-03},
	journal = {Physics Reports},
	author = {Kolodrubetz, Michael and Sels, Dries and Mehta, Pankaj and Polkovnikov, Anatoli},
	month = jun,
	year = {2017},
	pages = {1--87},
}

@String{Computer = "{IEEE} Computer" }

@String{Computing = "Computing" }

@STRING{IEEE = {Proc. {IEEE}}}

@STRING{MA = {Meteorological Applications}}

@String{Springer = "Springer-Verlag" }

@article{shaydulin2021classical,
  title={Classical symmetries and the Quantum Approximate Optimization Algorithm},
  author={Shaydulin, Ruslan and Hadfield, Stuart and Hogg, Tad and Safro, Ilya},
  journal={Quantum Information Processing},
  volume={20},
  number={11},
  pages={1--28},
  year={2021},
  publisher={Springer}
}

@article{kulshrestha2022beinit,
  title={BEINIT: Avoiding barren plateaus in variational quantum algorithms},
  author={Kulshrestha, Ankit and Safro, Ilya},
  journal={2022 IEEE International Conference on Quantum Computing and Engineering (QCE)},
  pages={197--203},
  year={2022},
  organization={IEEE}
}

@Article{shaydulin2018network,
  author    = {Shaydulin, Ruslan and Ushijima-Mwesigwa, Hayato and Safro, Ilya and Mniszewski, Susan and Alexeev, Yuri},
  journal   = {Advanced Quantum Technologies},
  title     = {Network community detection on small quantum computers},
  year      = {2019},
  number    = {9},
  pages     = {1900029},
  volume    = {2},
  publisher = {Wiley Online Library},
}

@INCOLLECTION{N,
  author = {Dontchev, A. L.},
  title = {Lipschitzian stability of {Newton}'s method for variational inclusions},
  booktitle = {System modeling and optimization (Cambridge, 1999)},
  publisher = {Kluwer Acad. Publ.},
  year = {2000},
  pages = {119--147},
  address = {Boston, MA}
}

@INCOLLECTION{AD,
  author = {Dontchev, A. L.},
  title = {Characterizations of {Lipschitz} stability in optimization},
  booktitle = {Recent Developments in Well-Posed Variational Problems},
  publisher = {Kluwer Acad. Publ.},
  year = {1995},
  chapter = {Math. Appl., 331},
  pages = {95--115}
}

@Article{tsvelikhovskiy2024symmetries,
  author  = {Boris Tsvelikhovskiy and Ilya Safro and Yuri Alexeev},
  title   = {Symmetries and Dimension Reduction in Quantum Approximate Optimization Algorithm},
  journal = {arXiv preprint arXiv:2309.13787},
  year    = {2026},
}

@article{tsvelikhovskiy2024equivariant,
  title={Equivariant QAOA and the Duel of the Mixers},
  author={Tsvelikhovskiy, Boris and Safro, Ilya and Alexeev, Yuri},
  journal={arXiv preprint arXiv:2405.07211},
  year={2024}
}

@Article{galda2023similarity,
  author   = {Galda, Alexey and Gupta, Eesh and Falla, Jose and Liu, Xiaoyuan and Lykov, Danylo and Alexeev, Yuri and Safro, Ilya},
  journal  = {Frontiers in Quantum Science and Technology},
  title    = {Similarity-based parameter transferability in the quantum approximate optimization algorithm},
  year     = {2023},
  issn     = {2813-2181},
  volume   = {2},
  doi      = {10.3389/frqst.2023.1200975},
  url      = {https://www.frontiersin.org/articles/10.3389/frqst.2023.1200975},
}

@article{egger2021warm,
  title={Warm-starting quantum optimization},
  author={Egger, Daniel J and Mare{\v{c}}ek, Jakub and Woerner, Stefan},
  journal={Quantum},
  volume={5},
  pages={479},
  year={2021},
  publisher={Verein zur F{\"o}rderung des Open Access Publizierens in den Quantenwissenschaften}
}

@article{herman2023quantum,
  title={Quantum computing for finance},
  author={Herman, Dylan and Googin, Cody and Liu, Xiaoyuan and Sun, Yue and Galda, Alexey and Safro, Ilya and Pistoia, Marco and Alexeev, Yuri},
  journal={Nature Reviews Physics},
  volume={5},
  number={8},
  pages={450--465},
  year={2023},
  publisher={Nature Publishing Group UK London}
}

@article{azad2022solving,
  title={Solving vehicle routing problem using quantum approximate optimization algorithm},
  author={Azad, Utkarsh and Behera, Bikash K and Ahmed, Emad A and Panigrahi, Prasanta K and Farouk, Ahmed},
  journal={IEEE Transactions on Intelligent Transportation Systems},
  year={2022},
  publisher={IEEE}
}

@INPROCEEDINGS{bach2024mlqaoa,
  author={Bach, Bao and Falla, Jose and Safro, Ilya},
  booktitle={2024 IEEE International Conference on Quantum Computing and Engineering (QCE)}, 
  title={MLQAOA: Graph Learning Accelerated Hybrid Quantum-Classical Multilevel QAOA}, 
  year={2024},
  volume={01},
  number={},
  pages={1-12},
  doi={10.1109/QCE60285.2024.00072}
}

@InProceedings{maciejewski2024multilevel,
  author       = {Maciejewski, Filip B and Bach, Bao G and Dupont, Maxime and Lott, P Aaron and Sundar, Bhuvanesh and Neira, David E Bernal and Safro, Ilya and Venturelli, Davide},
  booktitle    = {2024 IEEE High Performance Extreme Computing Conference (HPEC)},
  title        = {A multilevel approach for solving large-scale qubo problems with noisy hybrid quantum approximate optimization},
  organization = {IEEE},
  pages        = {1--10},
  year         = {2024},
}

@Article{falla2024graph,
  author    = {Falla, Jose and Langfitt, Quinn and Alexeev, Yuri and Safro, Ilya},
  title     = {Graph representation learning for parameter transferability in quantum approximate optimization algorithm},
  number    = {2},
  pages     = {46},
  volume    = {6},
  journal   = {Quantum Machine Intelligence},
  publisher = {Springer},
  year      = {2024},
}

@article{tate2023warm,
  title={Warm-Started QAOA with Custom Mixers Provably Converges and Computationally Beats Goemans-Williamson's Max-Cut at Low Circuit Depths},
  author={Tate, Reuben and Moondra, Jai and Gard, Bryan and Mohler, Greg and Gupta, Swati},
  journal={Quantum},
  volume={7},
  pages={1121},
  year={2023},
  publisher={Verein zur F{\"o}rderung des Open Access Publizierens in den Quantenwissenschaften}
}

@article{zhu2022adaptive,
  title={Adaptive quantum approximate optimization algorithm for solving combinatorial problems on a quantum computer},
  author={Zhu, Linghua and Tang, Ho Lun and Barron, George S and Calderon-Vargas, FA and Mayhall, Nicholas J and Barnes, Edwin and Economou, Sophia E},
  journal={Physical Review Research},
  volume={4},
  number={3},
  pages={033029},
  year={2022},
  publisher={APS}
}

@article{ushijima2021multilevel,
  title={Multilevel combinatorial optimization across quantum architectures},
  author={Ushijima-Mwesigwa, Hayato and Shaydulin, Ruslan and Negre, Christian FA and Mniszewski, Susan M and Alexeev, Yuri and Safro, Ilya},
  journal={ACM Transactions on Quantum Computing},
  volume={2},
  number={1},
  pages={1--29},
  year={2021},
  publisher={ACM New York, NY, USA}
}

@article{mcclean2018barren,
  title={Barren plateaus in quantum neural network training landscapes},
  author={McClean, Jarrod R and Boixo, Sergio and Smelyanskiy, Vadim N and Babbush, Ryan and Neven, Hartmut},
  journal={Nature communications},
  volume={9},
  number={1},
  pages={1--6},
  year={2018},
  publisher={Nature Publishing Group},
  doi={10.1038/s41467-018-07090-4},
  url={https://doi.org/10.1038/s41467-018-07090-4}
}

@InProceedings{tyagin2025qaoa,
  author       = {Tyagin, Ilya and Farag, Marwa H and Sherbert, Kyle and Shirali, Karunya and Alexeev, Yuri and Safro, Ilya},
  booktitle    = {2025 IEEE International Conference on Quantum Computing and Engineering (QCE)},
  title        = {{QAOA-GPT: Efficient generation of adaptive and regular quantum approximate optimization algorithm circuits}},
  organization = {IEEE},
  pages        = {1505--1515},
  volume       = {1},
  year         = {2025},
}

@Article{abbas2024challenges,
  author    = {Abbas, Amira and Ambainis, Andris and Augustino, Brandon and B{\"a}rtschi, Andreas and Buhrman, Harry and Coffrin, Carleton and Cortiana, Giorgio and Dunjko, Vedran and Egger, Daniel J and Elmegreen, Bruce G and others},
  title     = {Challenges and opportunities in quantum optimization},
  pages     = {1--18},
  journal   = {Nature Reviews Physics},
  publisher = {Nature Publishing Group},
  year      = {2024},
}

@InProceedings{nguyen2025cross,
  author       = {Nguyen, Kien X and Bach, Bao and Safro, Ilya},
  booktitle    = {{2025 IEEE International Conference on Quantum Computing and Engineering (QCE)}},
  title        = {Cross-Problem Parameter Transfer in Quantum Approximate Optimization Algorithm: A Machine Learning Approach},
  organization = {IEEE},
  pages        = {1--10},
  year         = {2025},
}

@Article{tsvelikhovskiy2026reductions,
  author  = {Tsvelikhovskiy, Boris and Bach, Bao and Falla, Jose and Safro, Ilya},
  title   = {Reductions of QAOA Induced by Classical Symmetries: Theoretical Insights and Practical Implications},
  journal = {arXiv preprint arXiv:2602.16141},
  year    = {2026},
}

\end{document}